\documentclass[graybox]{svmult}

\usepackage{mathptmx}       
\usepackage{helvet}         
\usepackage{courier}        
\usepackage{type1cm}        
\usepackage{makeidx}         
\usepackage{graphicx}        
\usepackage{multicol}        
\usepackage[bottom]{footmisc}

\usepackage{subfigure}
\usepackage{amsmath,amssymb,amsfonts}
\usepackage[format=plain,labelsep=quad,isu]{caption}
\usepackage{xfrac}

\def \im {\text{cor}}
\def \ex {\text{pre}}

\makeindex             

\begin{document}

\title*{Time-adaptive high-order compact finite difference schemes for option pricing in a family of stochastic volatility models}
\titlerunning{Time-adaptive high-order compact finite difference schemes}
  \author{Bertram D{\"u}ring and Christof Heuer}
\institute{Bertram D{\"u}ring \at Mathematics Institute, University of
  Warwick, Coventry, CV4 7AL, United Kingdom, \email{bertram.during@warwick.ac.uk}
\and Christof Heuer \at Frankfurt, Germany, \email{heuer.chr@googlemail.com}}
%
%
\maketitle

\abstract{We propose a {\em time-adaptive
  high-order compact
finite difference scheme\/} for option pricing in a {\em family of
stochastic volatility models}. 
We employ a semi-discrete high-order compact finite difference method
for the spatial discretisation, and combine this with an adaptive time discretisation, 
extending ideas from \cite{LSRH02} to fourth-order multistep methods in
time. }

\section{Introduction}
Stochastic volatility models have become
one of the standard approaches for financial option pricing. They are based on a two-dimensional stochastic diffusion process containing
two Brownian motions with correlation $\rho \in [-1,1]$, i.e.\
$\mathrm{E}\bigl[{\rm d}W_1(t) {\rm d}W_2(t)\bigr ]=\rho\, {\rm d}t,$ on a given filtered probability space for
the underlying asset $S=S(t)$ and the stochastic variance
$v=v(t)$. In this chapter we consider the following class of stochastic volatility models,
\begin{equation}
\label{General_SVmodel_SDE}
dS= \mu S\, {\rm d}t +\sqrt{v} S\,{\rm d}W_1,\quad
dv = \kappa v^a\left(\theta -v \right)\, {\rm d}t + \sigma v^b {\rm
  d}W_2 ,
\end{equation}
with given drift $\mu\in \mathbb{R}$ of the underlying $S(t)$, long run mean $\theta>0$, 
mean reversion speed $\kappa>0$, and volatility of volatility
$\sigma>0$, see e.g.\ \cite{ChJaMi08}. 
Additionally, it holds $a\geq 0$ and $b\in (0,3/2]$.
Many well-known models are included in the family \eqref{General_SVmodel_SDE}.
The prominent \textit{Heston (or SQR)
  model\/} \cite{Heston} is obtained for $a=0$, $b=1/2$.
Other known models include the \textit{GARCH (or VAR)
model\/} \cite{Duan95}, with $a=0$, $b=1$, and the \textit{3/2-model\/}
\cite{Lewis00} in which $a=0$, $b=3/2$. There are
also models with {\em non-linear\/} mean reversion,
following \cite{ChJaMi08}, we denote these models as the \textit{SQR-N
  model\/} ($a=1$, $b=1/2$), \textit{VAR-N model\/} ($a=1$, $b=1$),
and \textit{$3/2$-N model\/} ($a=1$, $b=3/2$).

For the family of stochastic volatility models \eqref{General_SVmodel_SDE},
application of It\^o's Lemma and standard arbitrage arguments lead to 
partial differential equations for the option price $V=V(S,v,t)$,
which are of the following form
\begin{equation}\label{pde_general_SV_model_untransformed}
\frac{\partial V}{\partial t} 
+ \frac{vS^2}{2}\frac{\partial^2 V}{\partial S^2} 
+ \rho \sigma v^{b+\frac{1}{2}} S\frac{\partial^2 V}{\partial S \partial v} 
+ \frac{\sigma^2 v^{2b}}{2}\frac{\partial^2 V}{\partial v^2}
 +rS\frac{\partial V}{\partial S} 
 + \kappa v^a \left(\theta -v\right) \frac{\partial V}{\partial v} 
 -rV =0,
\end{equation}
where $r\geq 0$
denotes the risk-free interest rate. Equation
\eqref{pde_general_SV_model_untransformed} has to be solved (backward
in time) for
$S,v>0,\,0 \leq t < T$, with an expiration date $T>0$, and subject to final and 
boundary conditions depending on the specific option considered. 
In the case of a European Put options, for example, the final condition is given by
$ V(S,v,T) = \max\left( K-S,0\right)$
with strike price $K>0$.


In the mathematical literature, there are many works on numerical
methods for option pricing in one-dimension (single
risk factor), but less papers considering numerical methods for option
pricing in stochastic volatility models, i.e.\ for two spatial
dimensions. Finite difference approaches used are
often standard, low order methods, i.e.\ second order in space. In the last decade, high-order (fourth order in space) compact finite
difference discretisations for option pricing in stochastic volatility models have been presented, e.g.\ in \cite{DuFo12,DuHe15}.
 We refer to
\cite{DuHe15} for an overview of the finite difference literature
and other methods.

The originality of the present chapter consists in proposing a new, {\em time-adaptive
  high-order compact
finite difference scheme\/} for option pricing in a {\em family of
stochastic volatility models}. 
Our approach builds on ideas from \cite{DuHe15} and \cite{LSRH02}.
We employ a semi-discrete high-order compact finite difference method
for the spatial discretisation, using the methodology developed in \cite{DuHe15}. 
For the adaptive time discretisation, 
we follow basic ideas of \cite{LSRH02}, where two-step methods for
the time-discretisation were used, and generalise this approach
to consider fourth-order multistep methods in time. We obtain a
time-adaptive high-order compact scheme that is fourth order accurate
in both space and time.

\section{Transformation of the partial differential equation}\label{sec:Transformations_SV_models}
We first transform
$\tau = T-t$, and $u = \exp(r\tau)V/K$ in \eqref{pde_general_SV_model_untransformed}.
Depending on the model parameter $b$, we apply subsequent
transformations, in such a way that the second derivatives in $x$- and
$y$-direction share the same coefficient.

For $b\neq {3}/{2}$  we apply the transformations
$x=( {3}/{2}-b) \ln( {{S}/{K}})$, $y={v}^{{3}/{2}-b}/{\sigma}$,
and arrive at
\begin{equation}\label{SVmodel_PDE_transformed_rectangular_shape_b_is_not_three_half}
u_\tau +a(y)\left(u_{xx} + u_{yy}\right)
+b(y)u_{xy} +c_1(y)u_x +c_2(y) u_y =0,
\end{equation}
to be solved on the rectangular spatial domain $\Omega=
(x_{\min},x_{\max})\times(y_{\min},y_{\max}),$
with 
\begin{align*}
\begin{split}
a(y) =& -{\sigma}^{{\frac {-5+2\,b}{-3+2\,b}}}{y}^{-2\, \left( -3+2\,b
 \right) ^{-1}} \left( -3+2\,b \right) ^{2}{(8\sigma)}^{-1},\quad b(y) =
2\rho a(y),\\
c_1(y)=&{( 3-2b)} \bigl( {\sigma}^{{\frac {-5+2\,b}{-3+2
\,b}}}{y}^{-2\, \left( -3+2\,b \right) ^{-1}}-2\,r\sigma \bigr) {
(4\sigma)}^{-1},\\
c_2(y) = &{ ( 3-2b)} \bigl( 2\,{\sigma}^{{\frac {-5+2\,b}{-3
+2\,b}}}{y}^{-{\frac {-1+2\,b}{-3+2\,b}}}b-4\,{\sigma}^{-{\frac {1+2\,
a-2\,b}{-3+2\,b}}}{y}^{-{\frac {1+2\,a-2\,b}{-3+2\,b}}}\kappa\,\theta 
\\
& +
4\,{\sigma}^{-{\frac {3+2\,a-2\,b}{-3+2\,b}}}{y}^{-{\frac {3+2\,a-2\,b
}{-3+2\,b}}}\kappa-{\sigma}^{{\frac {-5+2\,b}{-3+2\,b}}}{y}^{-{\frac {
-1+2\,b}{-3+2\,b}}} \bigr) {(8\sigma)}^{-1},
\end{split}
\end{align*}
and subject to 
 $u(x,y,0) = \max\left( 1-\exp\left({x}/({3}/{2}-b)\right),0\right).$

For $b=3/2$, we apply the transformations $x =\ln ({S}/{K})$, $y =\ln (v)/{\sigma}$,
and obtain \eqref{SVmodel_PDE_transformed_rectangular_shape_b_is_not_three_half}
with coefficients
$a(y) = - \exp(\sigma y)/2,$
$b(y) = - \rho \exp(\sigma y)$,
$c_1(y) =\exp(\sigma y)/2-r,$ 
$c_2(y)=(\sigma^{2}\exp(\sigma y)-2\kappa \theta\exp(\sigma y ( a-1)) +2\kappa\exp(a\sigma y))/(2\sigma)$
and subject to $u(x,y,0) = \max\left( 1-e^x,0\right).$

\section{Time-adaptive high-order compact scheme}
\label{sec:hoc}

We use the high-order compact semi-discrete
(discretising in space only) 
scheme from \cite{DuHe15} for
\eqref{SVmodel_PDE_transformed_rectangular_shape_b_is_not_three_half}. 
Since the coefficients of $u_{xx}$ and $u_{yy}$ in
\eqref{SVmodel_PDE_transformed_rectangular_shape_b_is_not_three_half}
are identical, results from \cite{DuHe15} show that
the scheme provides a fourth-order accurate spatial discretisation
employing a uniform grid with $h_1=h_2=h>0$. The semi-discrete scheme
can be written in matrix form as
\begin{align}\label{System_of_ODEs}
M_h\partial_\tau U_h\left(\tau\right) &=g^{(h)}\left(\tau\right) - K_h U^{(h)}\left(\tau\right) =:F\left(\tau\right).
\end{align}
The known vector $g^{(h)}$ has only non-zero entries due to the
influence of the boundary conditions and the matrices $M_h$ and $K_h$
do not depend on $\tau$.

At the boundary $x=x_{\min}$ and $x=x_{\max}$ we impose Dirichlet type
boundary conditions. For $y=y_{\min}$ or $y=y_{\max}$
we do not impose any boundary condition, but apply the discretisation
of the spatial interior. The resulting ghost points are extrapolated
from the interior with sufficiently high order.
Due to the low regularity of the typical initial conditions, we
employ a smoothing operator \cite{KrThWi70} to ensure fourth-order spatial
convergence.
For further details of the implementation of boundary and initial
conditions, we refer to \cite{DuHe15}.

Our approach for time adaptivity is motivated by \cite{LSRH02}, where
two-step methods are used for time discretisation. 
Here, to match the fourth-order accuracy in space, we consider
fourth-order multistep methods in time.
We approximate the system of ordinary differential
equations \eqref{System_of_ODEs} using fourth-order multistep methods
and variable, adaptive time step sizes. In each time step, we use a
(numerically cheap) \textit{predictor\/} scheme to estimate the local truncation
error, adapt the time step accordingly, and then solve using a
\textit{corrector\/} scheme.
Necessary start-up values are computed using a Crank-Nicolson time-discretisation.

\medskip
\noindent{\bf Predictor scheme.} 
Consider $\tau_{\min}=\tau_0<\tau_1<\ldots <\tau_{j}$ with $j\geq 4$
and $\tau_j< \tau_{\max}$ in time with the step sizes $k_n = \tau_n - \tau_{n-1} >0$
for $n=1,\ldots , j$. 
We denote the value of the vector $U^{(h)}$ at time $\tau_n$ by $U^{(h)}_{n}$. 

We use a four-step predictor scheme with (non-equidistant) time steps,
\begin{align}\label{Explicit_BDF_4}
        && \alpha_0^{(\ex)}M_h U^{(h)}_{n} =& k_n g^{(h)}_{n-1} -\left[ \alpha_1^{(\ex)} M_h + k_n K_h \right] U^{(h)}_{n-1} - M_h\sum\limits_{j=2}^4  \alpha_j^{(\ex)} U^{(h)}_{n-j} 
,
\end{align}
where
\begin{align*}
 \alpha_0^{(\ex)}=&
{ {
2\,\iota_{{1}}\iota_{{2}}\iota_{{3}}
+\iota_{{3}}{\iota_{{1}}^{2}}
+\iota_{{2}}{\iota_{{1}}^{2}}
+{\iota_{{2}}^{2}}\iota_{{3}}
+\iota_{{1}}{\iota_{{2}}^{2}}
}/{
\varphi_0^{(\ex)}
}},\\
\alpha_1^{(\ex)}=&
{\frac {
{\iota_{{1}}^{3}}\iota_{{2}}
-2\,\iota_{{1}}\iota_{{2}}\iota_{{3}}
-\iota_{{3}}{\iota_{{1}}^{2}}
-\iota_{{2}}{\iota_{{1}}^{2}}
-{\iota_{{2}}^{2}}\iota_{{3}}
-\iota_{{1}}{\iota_{{2}}^{2}}
+3\,{\iota_{{2}}^{2}}\iota_{{3}}\iota_{{1}}
+4\,{\iota_{{1}}^{2}}\iota_{{3}}\iota_{{2}}
+2\,{\iota_{{1}}^{2}}{\iota_{{2}}^{2}}
+{\iota_{{1}}^{3}}\iota_{{3}}
}{
2\,\iota_{{1}}\iota_{{2}}\iota_{{3}}
+\iota_{{3}}{\iota_{{1}}^{2}}
+\iota_{{2}}{\iota_{{1}}^{2}}
+{\iota_{{2}}^{2}}\iota_{{3}}
+\iota_{{1}}{\iota_{{2}}^{2}}
}},
\\
\alpha_2^{(\ex)}=&
-{ {
2\,\iota_{{1}}\iota_{{2}}\iota_{{3}}
+\iota_{{3}}{\iota_{{1}}^{2}}
+\iota_{{2}}{\iota_{{1}}^{2}}
+{\iota_{{2}}^{2}}\iota_{{3}}
+\iota_{{1}}{\iota_{{2}}^{2}}
}/{ 
\left( \left( 
\iota_{{2}}
+\iota_{{3}}
\right)  \left( 
\iota_{{1}}
+1 
\right) \right) 
}},
\\
\alpha_3^{(\ex)}=&
{ {
{\iota_{{2}}^{2}} 
\left( 
\iota_{{1}}\iota_{{2}}
+\iota_{{1}}\iota_{{3}}
+\iota_{{2}}\iota_{{3}} 
\right) 
}/{ 
\left( \left( 
\iota_{{1}}\iota_{{2}}
+\iota_{{2}}
+\iota_{{1}} 
\right)  \left( 
\iota_{{2}}
+\iota_{{1}} 
\right) \right) 
}},
\\
\alpha_4^{(\ex)}=&
{ { -
\left( 
\iota_{{2}}
+\iota_{{1}} 
\right) 
{\iota_{{2}}^{2}}{\iota_{{3}}^{4}}
}/{ 
\left( \left( 
\iota_{{1}}\iota_{{2}}\iota_{{3}}
+\iota_{{2}}\iota_{{3}}
+\iota_{{1}}\iota_{{3}}
+\iota_{{1}}\iota_{{2}}
 \right)  \left( 
\iota_{{1}}\iota_{{2}}
+\iota_{{1}}\iota_{{3}}
+\iota_{{2}}\iota_{{3}} 
\right)  \left( 
\iota_{{2}}
+\iota_{{3}}
 \right) \right) 
 }},
\end{align*}
with 
$\iota_{{1}}={k_n}/{k_{n-1}},$
$\iota_{{2}}={k_n}/{k_{n-2}}$, $\iota_{{3}} ={k_n}/{k_{n-3}} ,$
as well as
\begin{align*}
\varphi_0^{(\ex)} = & \text{ }
{\iota_{{1}}^{3}}\iota_{{3}}{\iota_{{2}}^{2}}
+3\,{\iota_{{2}}^{2}}\iota_{{3}}\iota_{{1}}
+4\,{\iota_{{1}}^{2}}\iota_{{3}}\iota_{{2}}
+2\,{\iota_{{1}}^{3}}\iota_{{3}}\iota_{{2}}
+3\,{\iota_{{1}}^{2}}{\iota_{{2}}^{2}}\iota_{{3}}
+{\iota_{{2}}^{2}}{\iota_{{1}}^{3}}
+2\,{\iota_{{1}}^{2}}{\iota_{{2}}^{2}}
\\
&
+{\iota_{{2}}^{2}}\iota_{{3}}
+2\,\iota_{{1}}\iota_{{2}}\iota_{{3}}
+\iota_{{1}}{\iota_{{2}}^{2}}
+\iota_{{3}}{\iota_{{1}}^{2}}
+{\iota_{{1}}^{3}}\iota_{{3}}
+{\iota_{{1}}^{3}}\iota_{{2}}
+\iota_{{2}}{\iota_{{1}}^{2}}.
\end{align*}
The predictor scheme \eqref{Explicit_BDF_4} is implicit. However,
since $M_h$ does not depend on $\tau$, it has to be factorised only
once at the beginning and the factorisation can then be re-used in
every time step. Hence, the predictor scheme is still computationally
cheap.

The local truncation error of the predictor scheme is given by
\begin{align}\label{local_error_explicit_four_step_method}
 \begin{split}
U^{(h)}\left(\tau_n\right) - \tilde{U}^{(h)}_n =&  C^{\text{loc}}_P k_n^5 \frac{\partial^5 u}{\partial \tau^5} + \mathcal{O}\left(k_n^6\right) ,
 \end{split}
\end{align}
with
$C_P^{\text{loc}}= 
 [ \left( \iota_{{1}}+1 \right)  \left( 
\iota_{{1}}\iota_{{2}}+\iota_{{2}}+\iota_{{1}} \right)  \left( 
\iota_{{1}}\iota_{{2}}\iota_{{3}}+\iota_{{2}}\iota_{{3}}+\iota_{
{1}}\iota_{{3}}+\iota_{{1}}\iota_{{2}} \right) ]/[120{{\iota_{{1}}}^{3}
\iota_{{3}}{\iota_{{2}}}^{2}}].$

In the following, we use the notation $\tilde{U}^{(h)}_n$ to clarify
whenever the predictor scheme is used to obtain the approximation of
the solution ${U}^{(h)}(\tau_n)$.

\medskip
\noindent{\bf Corrector scheme.}
For the corrector step, we use the implicit BDF-4 method with variable step-sizes to approximate the system of ordinary differential equations \eqref{System_of_ODEs},
\begin{align}\label{Implicit_BDF_4}
         &&\left[\alpha_0^{(\im)}M_h  + k_n K_h \right]U^{(h)}_{n}  =& - M_h\sum\limits_{j=1}^4  \alpha_j^{(\im)} U^{(h)}_{n-j}+ k_n g^{(h)}_n,
\end{align}
where 
\begin{align*}
 \alpha_0^{(\im)}=&
{\frac {
3\,{\iota_{{2}}^{2}}{\iota_{{1}}^{3}}
+4\,{\iota_{{1}}^{3}}\iota_{{3}}{\iota_{{2}}^{2}}
+6\,{\iota_{{1}}^{3}}\iota_{{3}}\iota_{{2}}
+2\,{\iota_{{1}}^{3}}\iota_{{2}}
+2\,{\iota_{{1}}^{3}}\iota_{{3}}
+9\,{\iota_{{1}}^{2}}{\iota_{{2}}^{2}}\iota_{{3}}
+4\,{\iota_{{1}}^{2}}{\iota_{{2}}^{2}}
+\iota_{{2}}{\iota_{{1}}^{2}}
}{ 
\left( 
\iota_{{1}}\iota_{{2}}\iota_{{3}}
+\iota_{{2}}\iota_{{3}}
+\iota_{{1}}\iota_{{3}}
+\iota_{{1}}\iota_{{2}} 
\right)  \left( 
\iota_{{1}}\iota_{{2}}
+\iota_{{2}}
+\iota_{{1}} 
\right)  \left( 
\iota_{{1}}+1 
\right)
}}
\\
&
     + {\frac {
     8\,{\iota_{{1}}^{2}}\iota_{{3}}\iota_{{2}}+
\iota_{{3}}{\iota_{{1}}^{2}}
+6\,{\iota_{{2}}^{2}}\iota_{{3}}\iota_{{1}}
+\iota_{{1}}{\iota_{{2}}^{2}}
+2\,\iota_{{1}}\iota_{{2}}\iota_{{3}}
+{\iota_{{2}}^{2}}\iota_{{3}}
}{ 
\left( 
\iota_{{1}}\iota_{{2}}\iota_{{3}}
+\iota_{{2}}\iota_{{3}}
+\iota_{{1}}\iota_{{3}}
+\iota_{{1}}\iota_{{2}} 
\right)  \left( 
\iota_{{1}}\iota_{{2}}
+\iota_{{2}}
+\iota_{{1}} 
\right)  \left( 
\iota_{{1}}+1 
\right)
}},
\\
\alpha_1^{(\im)} =&
-{\frac {
3\,{\iota_{{2}}^{2}}\iota_{{3}}\iota_{{1}}
+4\,{\iota_{{1}}^{2}}\iota_{{3}}\iota_{{2}}
+2\,{\iota_{{1}}^{3}}\iota_{{3}}\iota_{{2}}
+3\,{\iota_{{1}}^{2}}{\iota_{{2}}^{2}}\iota_{{3}}
+{\iota_{{1}}^{3}}\iota_{{3}}{\iota_{{2}}^{2}}
+{\iota_{{2}}^{2}}{\iota_{{1}}^{3}}
+2\,{\iota_{{1}}^{2}}{\iota_{{2}}^{2}}
+{\iota_{{2}}^{2}}\iota_{{3}}
}{ 
\left( 
\iota_{{1}}\iota_{{2}}
+\iota_{{1}}\iota_{{3}}
+\iota_{{2}}\iota_{{3}} 
\right)  \left( 
\iota_{{2}}
+\iota_{{1}} 
\right) 
}}
\\
&
-{\frac {2\,\iota_{{1}}\iota_{{2}}\iota_{{3}}+
\iota_{{1}}{\iota_{{2}}^{2}}
+\iota_{{3}}{\iota_{{1}}^{2}}
+{\iota_{{1}}^{3}}\iota_{{3}}
+{\iota_{{1}}^{3}}\iota_{{2}}
+\iota_{{2}}{\iota_{{1}}^{2}}
}{ 
\left( 
\iota_{{1}}\iota_{{2}}
+\iota_{{1}}\iota_{{3}}
+\iota_{{2}}\iota_{{3}} 
\right)  \left( 
\iota_{{2}}
+\iota_{{1}} 
\right) 
}},
\\
\alpha_2^{(\im)} =&
{\frac {
{\iota_{{1}}^{2}}{\iota_{{2}}^{2}}
+{\iota_{{1}}^{2}}{\iota_{{2}}^{2}}\iota_{{3}}
+2\,{\iota_{{1}}^{2}}\iota_{{3}}\iota_{{2}}
+\iota_{{2}}{\iota_{{1}}^{2}}
+\iota_{{3}}{\iota_{{1}}^{2}}
+2\,{\iota_{{2}}^{2}}\iota_{{3}}\iota_{{1}}
+\iota_{{1}}{\iota_{{2}}^{2}}
+2\,\iota_{{1}}\iota_{{2}}\iota_{{3}}
+{\iota_{{2}}^{2}}\iota_{{3}}
}{ 
\left( 
\iota_{{1}}
+1 
\right)  \left( 
\iota_{{2}}
+\iota_{{3}}
\right) 
}},
\\
\alpha_3^{(\im)} =&
-{\frac { 
\left( 
\iota_{{2}}\iota_{{3}}
+\iota_{{1}}\iota_{{3}}
+\iota_{{3}}{\iota_{{1}}^{2}}
+2\,\iota_{{1}}\iota_{{2}}\iota_{{3}}
+{\iota_{{1}}^{2}}\iota_{{3}}\iota_{{2}}
+\iota_{{2}}{\iota_{{1}}^{2}}
+\iota_{{1}}\iota_{{2}} 
\right) 
{\iota_{{2}}^{2}}
}{
\iota_{{2}}{\iota_{{1}}^{2}}
+\iota_{{1}}{\iota_{{2}}^{2}}
+2\,\iota_{{1}}\iota_{{2}}
+{\iota_{{2}}^{2}}
+{\iota_{{1}}^{2}}
}},
\\
\alpha_4^{(\im)} =&
{ {
\left( 
\iota_{{2}}
+\iota_{{1}}
+{\iota_{{1}}^{2}}
+2\,\iota_{{1}}\iota_{{2}}
+\iota_{{2}}{\iota_{{1}}^{2}} 
\right) 
{\iota_{{2}}^{2}}{\iota_{{3}}^{4}}
}/{
\varphi^{(\im)}_4
}},
\end{align*}
with  $\iota_{{1}}={k_n}/{k_{n-1}},$
$\iota_{{2}}={k_n}/{k_{n-2}}$, $\iota_{{3}} ={k_n}/{k_{n-3}} ,$
as well as
\begin{align*}
\varphi^{(\im)}_4 =&
\text{ }
{\iota_{{3}}^{3}}{\iota_{{1}}^{2}}
+2\,\iota_{{1}}{\iota_{{3}}^{3}}\iota_{{2}}
+4\,\iota_{{1}}{\iota_{{2}}^{2}}{\iota_{{3}}^{2}}
+{\iota_{{2}}^{2}}{\iota_{{3}}^{3}}
+2\,{\iota_{{1}}^{2}}{\iota_{{2}}^{2}}{\iota_{{3}}^{2}}
+{\iota_{{2}}^{2}}{\iota_{{3}}^{3}}\iota_{{1}}
+{\iota_{{2}}^{3}}{\iota_{{3}}^{2}}
\\
&
+{\iota_{{1}}^{2}}{\iota_{{2}}^{3}}
+\iota_{{2}}{\iota_{{3}}^{3}}{\iota_{{1}}^{2}}
+{\iota_{{2}}^{3}}{\iota_{{3}}^{2}}\iota_{{1}}
+3\,{\iota_{{1}}^{2}}\iota_{{2}}{\iota_{{3}}^{2}}
+3\,{\iota_{{1}}^{2}}{\iota_{{2}}^{2}}\iota_{{3}}
+{\iota_{{2}}^{3}}\iota_{{3}}{\iota_{{1}}^{2}}
+2\,\iota_{{1}}{\iota_{{2}}^{3}}\iota_{{3}}.
\end{align*}
The local truncation error of the corrector scheme is given by
\begin{align}\label{local_error_BDF4_implicit}
 \begin{split}
U^{(h)}\left(\tau_n\right) - U^{(h)}_n =&  C^{\text{loc}}_C k_n^5 \frac{\partial^5 U^{(h)}(\tau_n)}{\partial \tau^5} + \mathcal{O}\left(k_n^6\right) ,
 \end{split}
\end{align}
with
\begin{align*}
C^{\text{loc}}_C =&  \,{{ 
-\left( \iota_{{1}}\iota_{{2}}\iota_{{3}
}+\iota_{{2}}\iota_{{3}}+\iota_{{1}}\iota_{{3}}+\iota_{{1}}\iota
_{{2}} \right) ^{2}
\left( \iota_{{1}}+1 \right) ^{2} 
\left( \iota_{{1}}\iota_{{2}}+\iota_{{2}}+\iota_{{1}} \right) ^{2}
}/{
N_C^{\text{loc}}
}},
\\
N_C^{\text{loc}}=&
120
{\iota_{{1}}^{3}}{\iota_{{2}}^{2}}\iota_{{3}} 
\left[ 
4\,{\iota_{{1}}^{3}}\iota_{{3}}{\iota_{{2}}^{2}}
+6\,\iota_{{3}}{\iota_{{1}}^{3}}\iota_{{2}}
+2\,\iota_{{3}}{\iota_{{1}}^{3}}
+3\,{\iota_{{2}}^{2}}{\iota_{{1}}^{3}}
+2\,\iota_{{2}}{\iota_{{1}}^{3}}
+8\,{\iota_{{1}}^{2}}\iota_{{3}}\iota_{{2}}
\right.
\\
&\left.
+9\,{\iota_{{1}}^{2}}{\iota_{{2}}^{2}}\iota_{{3}}
+{\iota_{{1}}^{2}}\iota_{{3}}
+4\,{\iota_{{1}}^{2}}{\iota_{{2}}^{2}}
+\iota_{{2}}{\iota_{{1}}^{2}}
+6\,{\iota_{{2}}^{2}}\iota_{{3}}\iota_{{1}}
+2\,\iota_{{1}}\iota_{{2}}\iota_{{3}}
+\iota_{{1}}{\iota_{{2}}^{2}}
+\iota_{{3}}{\iota_{{2}}^{2}} \right] .
\end{align*}

\medskip
\noindent{\bf Time-step adaption.}
The aim of the time-step adaption is to choose the time-step in such a
way that the resulting local time-discretisation error stays below a
given threshold $\hat{\epsilon}>0$.
Similar as in \cite{PeSy07},
we use the local time-discretisation errors \eqref{local_error_explicit_four_step_method} and \eqref{local_error_BDF4_implicit} to obtain the first order approximation
\begin{align}\label{approximation_fifth_derivative_in_time_with_explicit_and_implicit_methods}
\frac{\partial^5 U^{(h)}\left(\tau_n\right)}{\partial \tau^5} =&
\frac{U_n^h - \tilde{U}_n^h}{k_n^5\left(C^{\text{loc}}_C - C^{\text{loc}}_P\right)} + \mathcal{O}\left( k_n \right) .
\end{align}
The leading error term of the discretisation \eqref{Implicit_BDF_4} can thus be approximated by
\begin{align}\label{leadingerrorapproximation}
 \epsilon_n = &
 -\alpha_0^{(\im)}M_h C_C^{(\text{loc})}k_n^4 \frac{\partial^5 U^{(h)}}{\partial \tau^5}
=-\alpha_0^{(\im)}
M_h C_C^{(\text{loc})} \frac{U_n^h - \tilde{U}_n^h}{k_n\left(C^{\text{loc}}_C - C^{\text{loc}}_P\right)} .
\end{align}
The goal is now to choose the next step-size in time in a way that the
norm of this error is bounded by the error threshold $\hat{\epsilon}>0$ in a given norm.
The general error structure is given by
$ \epsilon_n  = k_n^4 \zeta (\tau_n)\;\Longleftrightarrow \; k_n = ({\epsilon_n}/{\zeta (\tau_n)})^{\frac{1}{4}} $
and thus we can, with $\Vert \epsilon_n\Vert\leq \hat{\epsilon}$, use
$k_{n+1} \leq k_n({\hat{\epsilon}}/{\left\Vert
      \epsilon_n\right\Vert })^\frac{1}{4}$ to choose the new step size in time.

The approximation of the local discretisation error in time
\eqref{leadingerrorapproximation} can be non-smooth, giving
rise to abrupt changes of the chosen step size.
To ensure that we avoid choosing a very large step size in case that
the estimated error is very small, we introduce a small parameter
$\beta>0$ (see \cite{PeSy07}) and adapt the time step size according to
\begin{align}\label{definition_multiplicator_xi}
 k_{n+1} 
=\left(\frac{\hat{\epsilon}}{\hat{\epsilon} \beta + \left\Vert {\epsilon}_n\right\Vert }\right)^\frac{1}{4}k_n=:\xi_n k_n.
\end{align}


\section{Numerical results}
\label{sec:numeric}

We consider the pricing of European Put options with model
\eqref{General_SVmodel_SDE} and use
$(S,v)\in (1.5,600)\times (0.1,0.5)$.
The computational domain is determined through the transformations given in Section~\ref{sec:Transformations_SV_models}.
We choose step-size $h=(x_{\max} - x_{\min})/{(N-1)}$ with 
$N=201$ steps in $x$-direction, in $y$-direction we begin
at $y_{\min}$ and use step-size $h$. In \eqref{definition_multiplicator_xi}, we
set $\beta=0.01$. 
We use $K=100$,
 $T=2$,
$r=0.05$,
  $\sigma=0.3$, 
  $\kappa =1.1$,
$\theta =0.3$,
$\rho=-0.4$.
For the start-up values, we apply the Crank-Nicolson time-steps with a
fixed parabolic mesh ratio, choosing  $k_n= 0.05h^2,$ $n=1,2,3$. 

\begin{figure}
  \begin{minipage}[t]{.49\textwidth}
    \centering
    \includegraphics[width=\textwidth]{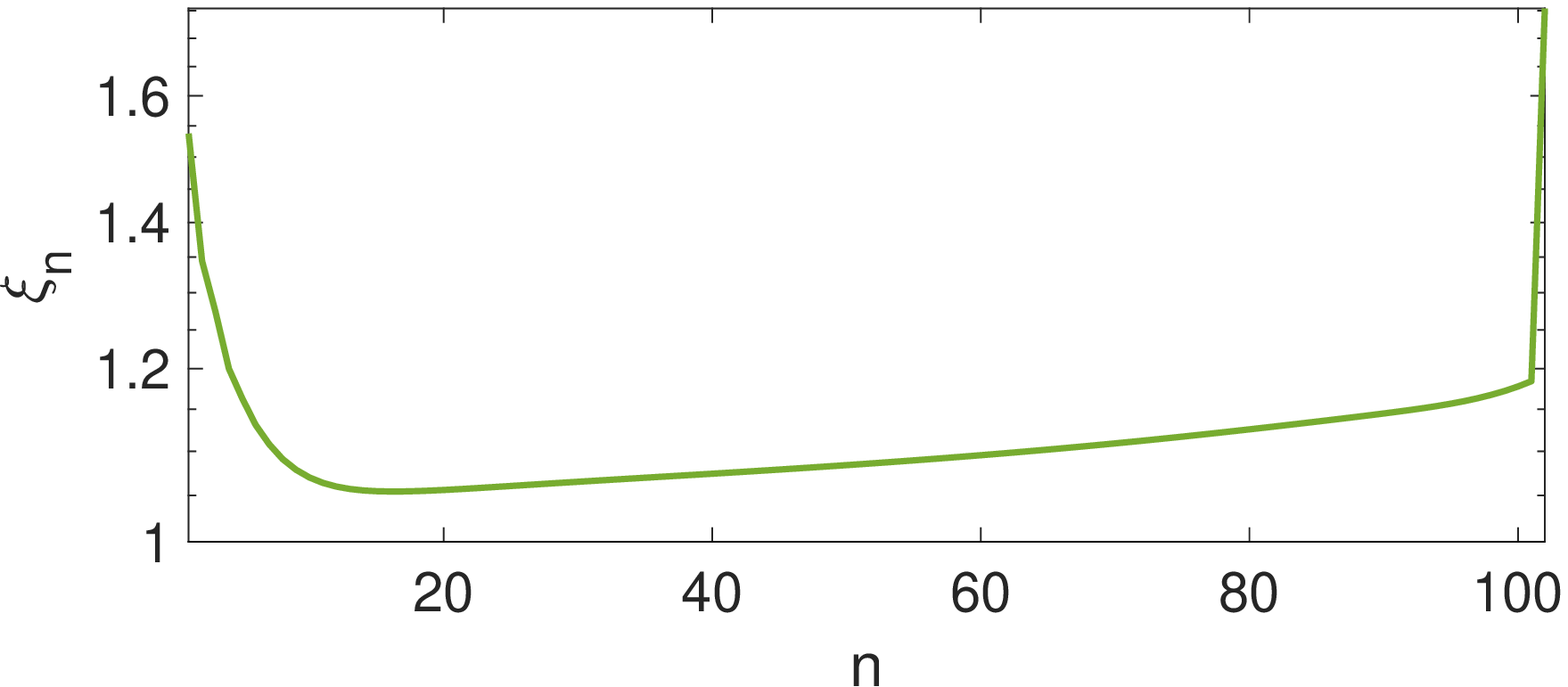}\\[0.2em]
    \hfill\includegraphics[width=0.9\textwidth]{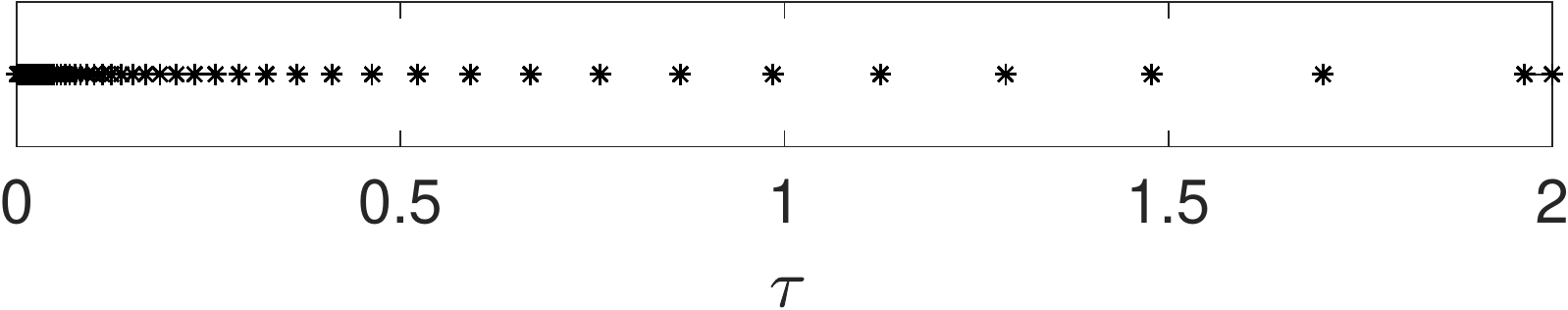}
    \includegraphics[width=\textwidth]{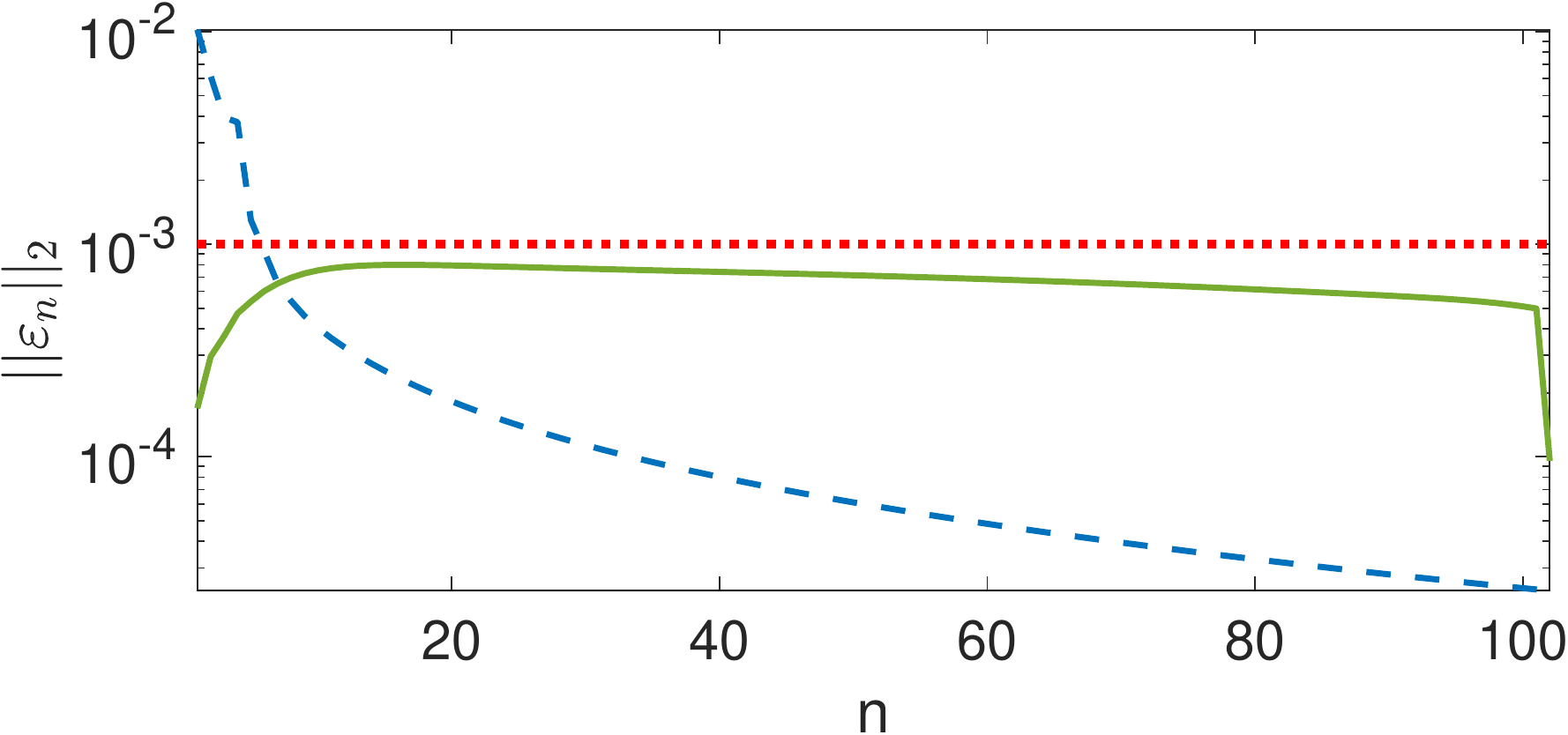}
  \end{minipage}%
  \hspace*{0.02\textwidth}
  \begin{minipage}[t]{.49\textwidth}
    \centering
    \includegraphics[width=\textwidth]{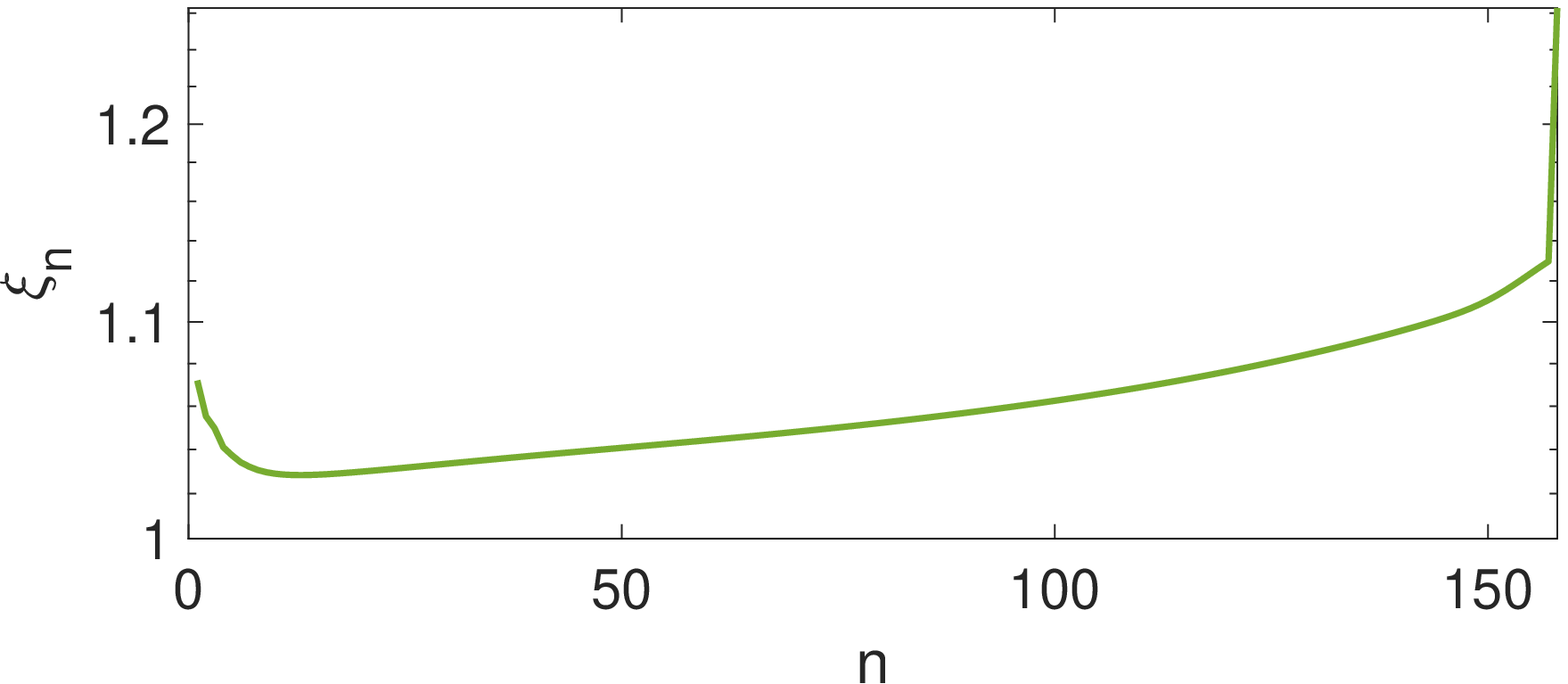}\\[0.2em]
    \hfill\includegraphics[width=0.9\textwidth]{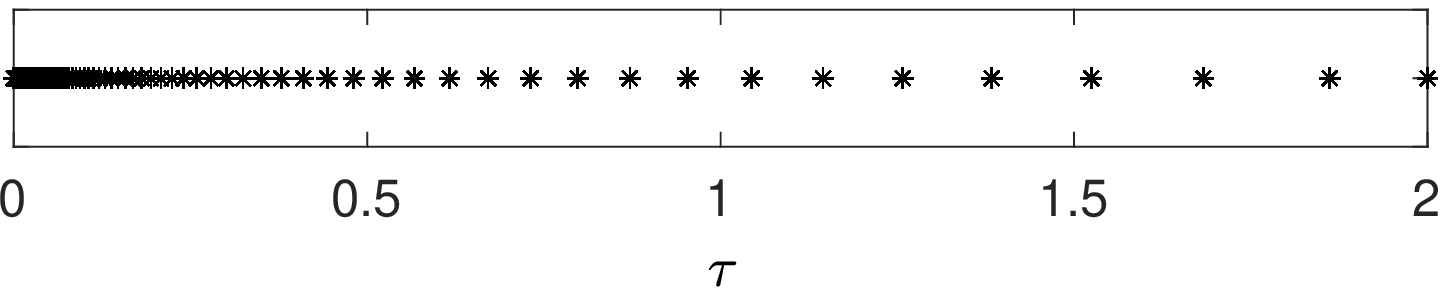}\\[0.3em]
    \includegraphics[width=\textwidth]
    {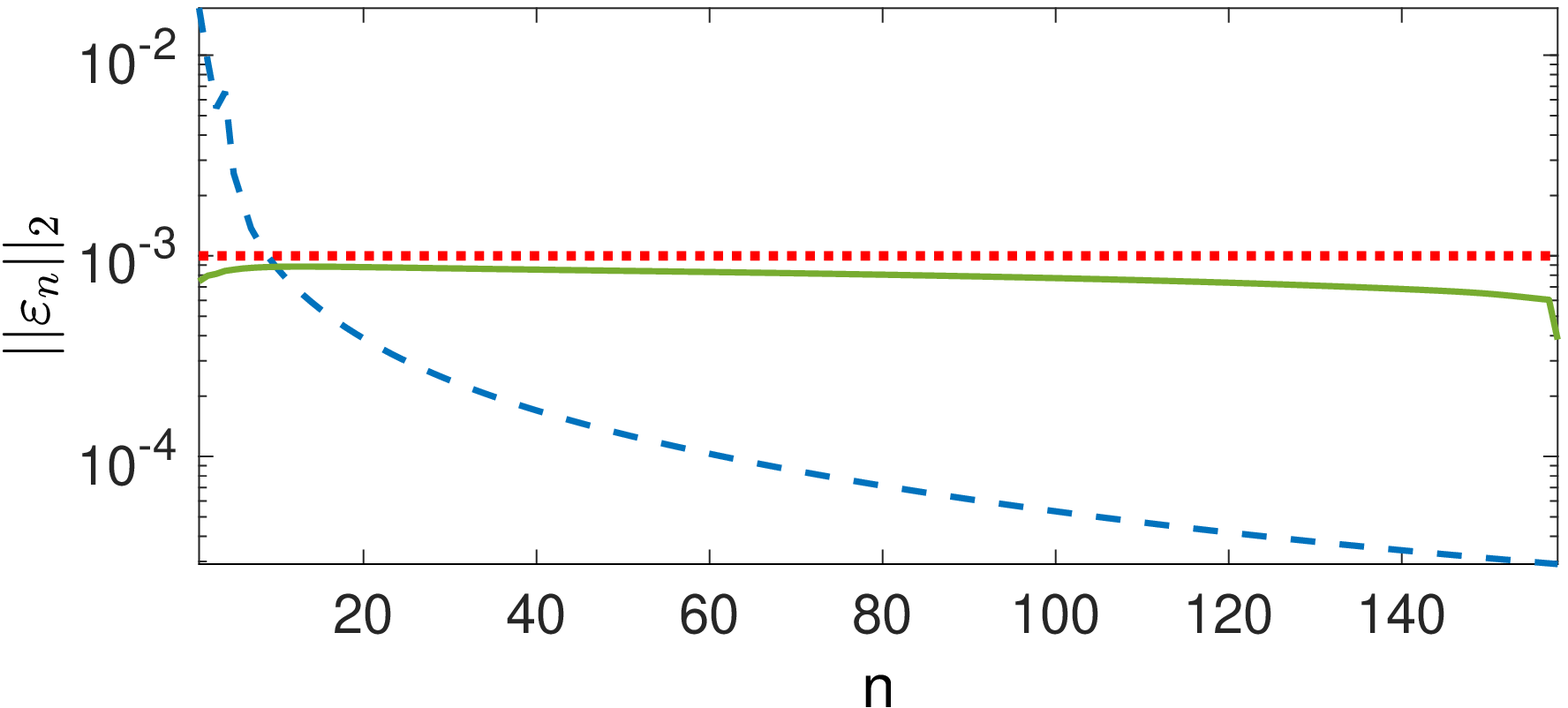}      
  \end{minipage}%
  \caption{Adaptation factor $\xi_n$, time grid points distribution, and
    error threshold $\hat{\epsilon}$ (dotted red), local error $|| \epsilon_n||_2$ for adaptive (solid green) and equidistant time
    stepping (dashed blue): GARCH (left), $a=b=3/4$ model
    (right).}
  \label{fig:resultsGARCH34}
\end{figure}

Figure~\ref{fig:resultsGARCH34} shows the adaptation factor $\xi_n$, the
positioning of the grid points in time, and the local error
$||\epsilon_n||_2$ for the GARCH model (left column) and the $a=b=3/4$
model (right column).
For GARCH the algorithm leads to overall $104$ grid-points in time. The local error
remains just below the chosen threshold $\hat{\epsilon}=0.001$, while time steps are
increased. For GARCH, $50$ of $104$ grid-points in time, including the
three initial points where Crank-Nicolson type time discretisation is
used, are located in the interval $[0,0.01]$, i.e.\ $48\%$ of the
grid-points are positioned in only $0.5\%$ of the time-domain. On the
other hand only six points are placed in the time interval
$[1,2]$. The results for the $a=b=3/4$ model show a similar behaviour.
For comparison we repeat both simulations, now with the same numbers of
{\em equidistant\/} time steps. Initially, the local error
is above the threshold and later far below, indicating the
sub-optimality of the equidistant distribution of points in time.
  

\bibliographystyle{alpha} 
\bibliography{referenc_duering}

\end{document}